\documentclass{PoS}
\usepackage{amsmath}
\usepackage{amsfonts}
\usepackage{amssymb}
\usepackage{fontenc}
\usepackage{times}
\usepackage{mathptmx}
\usepackage{graphicx}
\usepackage{graphics,epsfig}

\title{Implications of local parity breaking in heavy ion collisions}

\ShortTitle{$e^+e-$ excess from LPB}

\author{\speaker{Xumeu Planells}\\
Departament d'Estructura i Constituents de la Matèria and Institut de Ciències del Cosmos (ICCUB) , Universitat de Barcelona, Spain\\
E-mail: \email{xumeu@icc.ub.edu}}
\author{Alexander A. Andrianov and Vladimir A. Andrianov
\\ V. A. Fock Department of Theoretical Physics, Saint-Petersburg State University, Russia\\
E-mail: \email{sashaandrianov@gmail.com}, \email{v.andriano@rambler.ru}}
\author{Domenec Espriu\\
Departament d'Estructura i Constituents de la Matèria and Institut de Ciències del Cosmos (ICCUB) , Universitat de Barcelona, Spain\\
E-mail: \email{espriu@ecm.ub.es}}

\abstract{Recent data on dilepton production in heavy ion collisions revealed an abnormal excess in the region of invariant masses below 1 GeV. Our proposal is the creation of a slowly varying time-dependent pseudoscalar condensate within the hot nuclear fireball that comes from the very collision. The local parity breaking effect that immediately arises substantially modifies the vector meson properties leading to an excess of lepton pairs that could be a part of the explanation for the observed abnormal dilepton yield.}

\FullConference{The XXth International Workshop High Energy Physics and Quantum Field Theory\\
		September 24-October 1, 2011\\
		Sochi, Russia}

\begin{document}

\section{Motivation of local parity breaking}

It is well known that parity is a well established global symmetry of strong interactions. However, some time ago, it was proposed that the QCD vacuum can possess metastable domains leading to $P$ violation. Accordingly, there is no reason to think that this symmetry cannot be broken in a finite volume. It is conjectured that the presence of non-trivial angular momentum (or magnetic field) in heavy ion collisions (HIC) then leads to the so-called Chiral Magnetic Effect (CME) \cite{cme}.

\begin{figure}[h!]
\begin{minipage}[t]{7.35cm}



\qquad Last decade several experiments in central HIC have indicated an abnormal yield of $e^+e^-$ pairs of invariant mass < 1 GeV. This is the case of the PHENIX experiment in Brookhaven National Laboratory's Relativistic Heavy Ion Collider (RHIC) (see Fig. \ref{phen}). It has been well established that such an enhancement is certainly a nuclear medium effect when compared to proton-proton collisions. The abnormal dilepton yield (enhanced by a factor of $\sim 4.7$) has not been yet explained by any of the available mechanisms in hadron phenomenology. In this context, we want to stress that both CME and dilepton excess may be complementary effects related to the formation of a new phase in QCD 



\end{minipage}\qquad \begin{minipage}[t]{7cm}
\centering
\vspace{-0.5em}\includegraphics[scale=0.25]{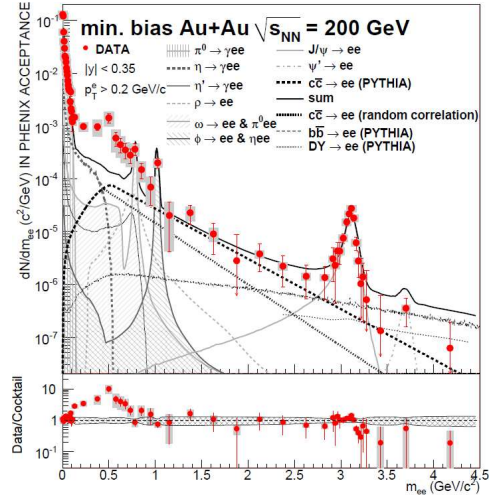}
\vspace{-1em}
\caption{Anomalous dilepton yield in Au+Au collisions in PHENIX as compared with p+p \cite{phenix}.\label{phen}}

\end{minipage}
\end{figure}

\vspace{-1em}\hspace{-2em}where parity is locally broken. Our goal, with all the machinery developed for the study of local parity breaking (LPB) in HIC, consists in giving a qualitative answer to the PHENIX anomaly.


\section{Axial baryon charge and chiral chemical potential}

The strong interaction shows a highly non-trivial vacuum energy due to its non-abelian character. Such a behaviour allows different topological sectors to be present in the vacuum state, separated by high energy barriers. Nevertheless, these non-equivalent configurations may be connected due to large quantum fluctuations of the vacuum state in the presence of a hot medium via sphaleron transitions. The topological charge $T_5$ in a finite volume associated with a fireball reads
\begin{equation}
T_5=\frac{1}{8\pi^2}\int_{\text{vol.}}d^3x~ \varepsilon_{jkl}\text{Tr}\left (G^j\partial^k G^l-i\frac23G^jG^kG^l\right ),
\end{equation}
which is not a gauge invariant object under large gauge transformations. However, its jump can be associated to the integral of a gauge-invariant quantity. Then, in order to examine LPB in nuclear matter, we can integrate the local PCAC and connect $T_5$ and the quark axial charge $Q_5^q$:
\begin{equation}\label{pcac}
Q_5^q=\int_{\text{vol.}}d^3x~\bar q\gamma_0\gamma_5q,\qquad
\frac d{dt}\left (Q_5^q-2N_f T_5\right )\simeq 0,~m_q\simeq 0.
\end{equation}
In the chiral limit, thus, the exact conservation of the axial current states that the presence of a non-trivial topological charge in a finite volume is directly related to the emergence of a quark axial charge, a clear indication LPB.


The incorporation of QED into this framework entails the addition of an extra electromagnetic axial charge $Q_5^{\text{em}}$. At this point, the sum of both contributions is the one directly related to the change of $T_5$ in Eq. \eqref{pcac}. In this sense, the description of the total axial charge may be described by a chiral chemical potential $\mu_5$.

\vspace{0.5em}

After bosonization of QCD in the light meson sector following the standard Vector Meson Dominance (VMD) prescription, the quark axial charge $Q_5^q$ is substituted by its equivalent operator incorporating the lightest vector mesons fields. Accordingly, the QED Lagrangian may be extended with an extra piece given by the anomalous term $\Delta\mathcal L\sim -\frac14\varepsilon^{\mu\nu\rho\sigma}\text{Tr}\left [\hat\zeta_\mu V_\nu V_{\rho\sigma}\right ],$ where $\hat\zeta_\mu$ has to be understood as the derivative of some time dependent but approximately spatially homogeneous background pseudoscalar field $a(t)$ induced within the fireball resulting from a HIC. Therefore, we are just left with one component of $\hat \zeta_\mu\simeq\partial_\mu\hat a\simeq \hat \zeta\delta_{\mu 0}$, where the hat notation stands for the isospin content that carried by this object. In addition, the absolute magnitude of $\hat \zeta$ is directly associated to the chiral chemical potential $\zeta\sim \alpha\mu_5$ and one may expect it to be of order $\zeta\sim\alpha \tau^{-1}\sim 1$ MeV, taking the condensate time of formation $\tau\sim1$ fm and the value of condensate of order of $f_\pi$.

\vspace{0.5em}

In summary, the existence of a non-trivial thermal average of the topological charge is translated into a prediction of $\zeta$. However, we propose the inverse procedure: \emph{if one is able to measure the average value of $\zeta$ in a HIC, then the topological charge average can be automatically found using known techniques on the lattice.} In this work, we provide some reasonable values for $\langle\zeta\rangle$ in our attempt to unravel the anomalous
dilepton production in the PHENIX experiment.


\section{Vector Meson Dominance approach to LPB}

The appropriate framework to describe electromagnetic interactions of hadrons at low energies is the VMD containing the lightest vector mesons $\rho_0$, $\omega$ and $\phi$ in the $SU(3)$ flavor sector. Quark-meson interactions are described by
\begin{equation}
{\cal L}_{\text{int}} = \bar q \gamma_\mu \hat V^\mu q;\quad \hat V_\mu \equiv - e A_\mu Q  + \frac12 g_\omega  \omega_\mu \textbf{I}_{ns} + \frac12g_\rho \rho_\mu^0  \tau_3+\frac1{\sqrt 2}g_\phi\phi_\mu\textbf I_s,
\end{equation}
where $(V_{\mu,a}) \equiv \left(A_\mu,\, \omega_\mu, \, \rho_\mu^0, \phi_\mu\right)$, $Q=\frac{1}2\tau_3+\frac16\textbf{I}_{ns}-\frac13\textbf{I}_s$, and $g_\omega \simeq  g_\rho \equiv g \simeq 6<g_\phi\simeq 7.8$. These values are extracted from vector meson decays. The Maxwell and mass terms are
\begin{gather}
{\cal L}_{\text{kin}} = - \frac14 \left(F_{\mu\nu}F^{\mu\nu}+ \omega_{\mu\nu}\omega^{\mu\nu}+ \rho_{\mu\nu}\rho^{\mu\nu}+\phi_{\mu\nu}\phi^{\mu\nu}\right) + \frac12  V_{\mu,a}(\hat m^2)_{ab}V^\mu_{b}
\end{gather}
\begin{gather}
\hat m^2\simeq m_V^2\left(\begin{array}{ccccccc}
\frac{4 e^2}{3g^2} & &-\frac{e}{3g} && -\frac{e}{g} && \frac{\sqrt 2eg_\phi}{3g^2}\\
 -\frac{e}{3g}&& 1 && 0 && 0\\
 -\frac{e}{g} && 0 && 1 && 0\\
\frac{\sqrt 2eg_\phi}{3g^2} && 0 && 0 && \frac{g_\phi^2}{g^2} \\
\end{array}\right),
\end{gather}
where the mixing among vector mesons and photons appears. In addition, as stressed before, a parity-odd term has to be taken into account that produces an extra mixing:

\begin{equation}
{\cal L}_{\text{mix}}=\frac12 \text{Tr}\left (\hat\zeta \varepsilon_{jkl} \hat V_{j} \partial_k \hat V_{l} \right ) = \frac12 \zeta \varepsilon_{jkl} V_{j,a} N_{ab}\partial_k V_{l,b},
\end{equation}

In a HIC, where a fireball of a lifetime $\tau_{\text{fb}}\sim 5-10$ fm is created, the chiral charge that arises due to a jump in the topological charge has a characteristic oscillation time governed by inverse quark masses. Thus, for $u$, $d$ quarks $1/m_q\sim1/(5$ MeV$)\sim 40$ fm$\gg\tau_{\text{fb}}$, so the oscillation can be neglected and during the fireball lifetime we may consider the chiral charge to be approximately constant. However, for the $s$ quark the situation is different as $1/m_s\sim1/(200$ MeV$)\sim 1$ fm$\ll\tau_{\text{fb}}$ and even that a topological charge persists during fireball lifetime, the mean value of strange quark chiral charge is around zero due to essential left-right oscillations. Furthermore, the $\phi$ meson lifetime is much larger than the fireball; so finally one expects the strange sector to decouple from the rest of the system, thus being a good approximation to reduce from 3 to 2 flavors. In this sense, from now on we shall just consider a general $\zeta$ to be a linear combination of isosinglet and isotriplet cases, in the same way as in \cite{pos}.

\vspace{0.5em}

In the case of an experiment that creates a very hot nuclear fireball rather than dense, \textit{i.e.} $T\gg\mu$; an isosinglet pseudoscalar background is expected to be dominant, as it is the case for RHIC and LHC. In such environment $e^2\hat\zeta=9/5\zeta\textbf{I}$, with a mixing matrix $N^\theta$. Unlikely, when the opposite condition $T\ll\mu$ holds, like for the future experiments FAIR and NICA, the expected dominant background is the isotriplet one, where $e^2\hat\zeta=3\zeta\tau_3$, with a CS matrix $N^\pi$:
\begin{equation}
(N_{ab}^\theta)\, \simeq\,  \left(\begin{array}{ccccc}
1 &  &-\frac{3g}{10e}& &-\frac{9g}{10e}\\
-\frac{3g}{10e}& & \frac{9g^2}{10e^2} & & 0 \\
-\frac{9g}{10e}& & 0 & &\frac{9g^2}{10e^2} \\
\end{array}
\right)\sim (m_V^2)_{ab}\Bigg |_{SU(2)_f},\quad (N_{ab}^\pi)\, \simeq\,  \left(\begin{array}{ccccc}
1 &  &-\frac{3g}{2e}& &-\frac{g}{2e}\\
-\frac{3g}{2e}& & 0 & & \frac{3g^2}{2e^2} \\
-\frac{g}{2e}& & \frac{3g^2}{2e^2} & & 0 \\
\end{array}
\right).
\end{equation}
In this work, we shall not consider about the second case and we will just focus on the first one that could be a source of dilepton enhancement to give a qualitative answer to the PHENIX anomaly. Following these lines, one may solve the equations of motion after simultaneous diagonalization of the matrices $m^2$ and $N^\theta$. Then, it is possible to show that photons remain massless and undistorted while mesons split into three polarizations ($\epsilon=0,\pm1$) with masses
\begin{equation}
m_{V,\epsilon}^2=m_V^2-\epsilon\frac{9g^2}{10e^2}\zeta|\vec k|\simeq m_V^2-360~\epsilon\zeta|\vec k|.
\end{equation}
This splitting unambiguously signifies local parity breaking as well as breaking of Lorentz invariance due to the time-dependent background. The enlargement of the resonant region potentially leads to a substantial enhancement of their contribution to dilepton production away from their nominal vacuum resonance position.

\vspace{0.5em}

On the other side, the averaged topological charge should be expected to vanish due to an equal probability to find positive and negative jumps. However, the change in the sign of $\zeta$ is equivalent to the change of the $+$ polarization to the $-$ one, producing the same result. As a consequence, we shall just focus in the averaged modulus of $\zeta$.





\section{Manifestation of LPB in heavy ion collisions}
\begin{figure}[h!]
\centering
\includegraphics[scale=0.2]{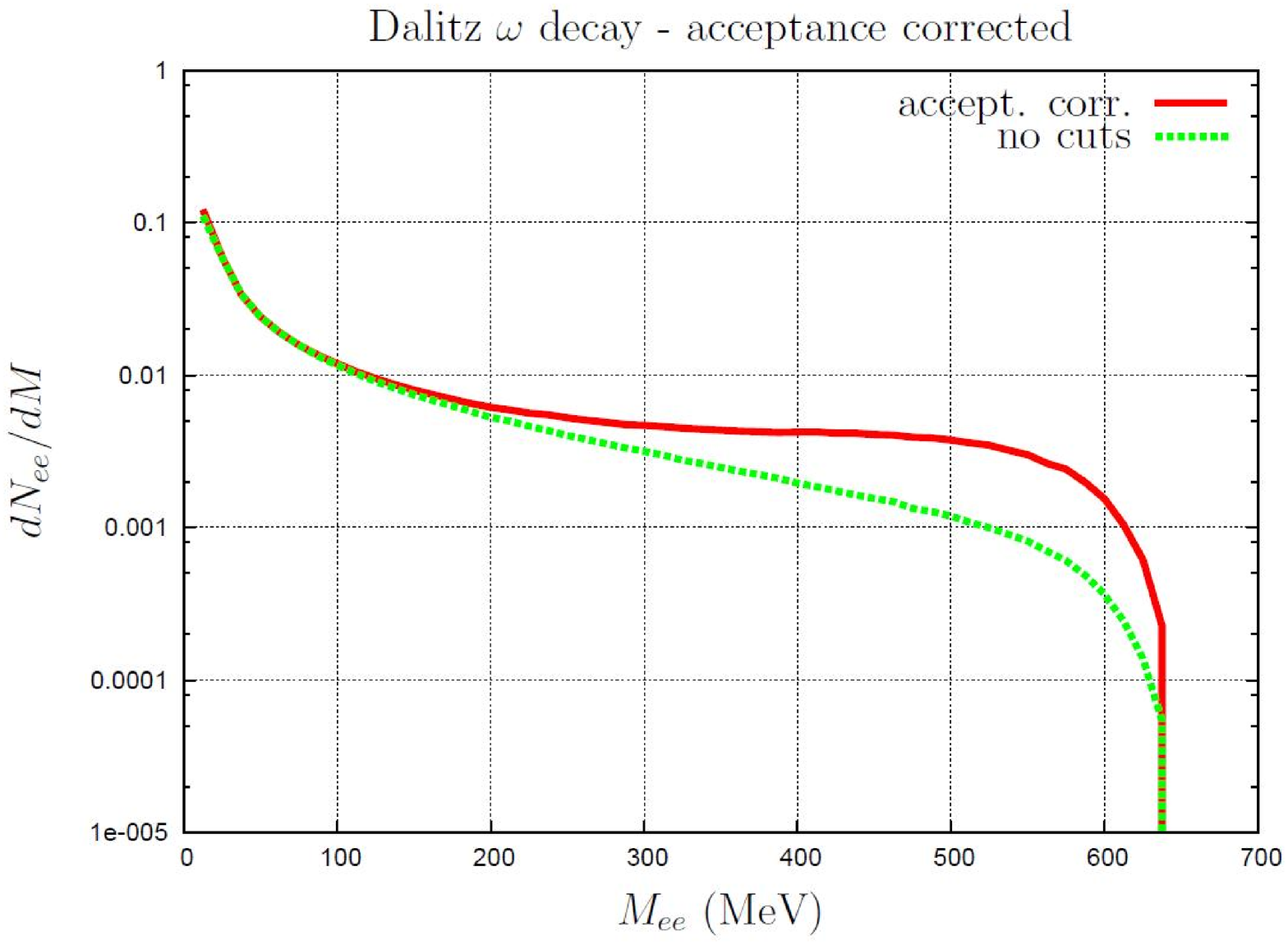}\quad\includegraphics[scale=0.2]{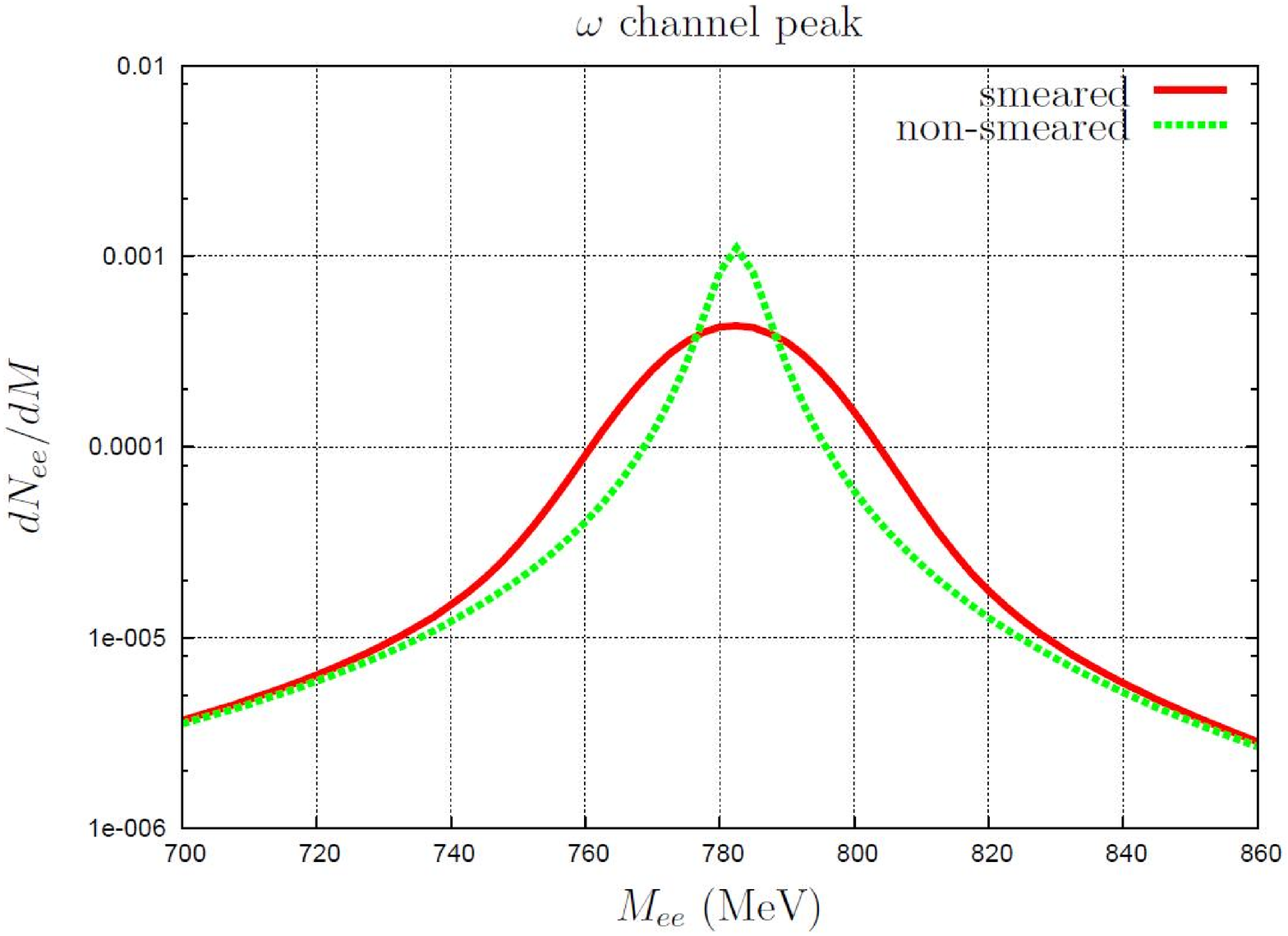}
\vspace{-1em}
\caption{Acceptance correction. Left: Dalitz $\omega$ decay with (and without) experimental cuts. Right: Detail of the $\omega$ peak taking into account mass smearing.}\label{accept}
\end{figure}

The main contributions in the cocktail of hadron decays presented by the PHENIX collaboration in the intermediate mass region $300<M_{ee}<900$ MeV comes from the direct processes $\rho,\omega\rightarrow e^+e^-$, the Dalitz decays $\pi^0,\eta,\eta'\rightarrow\gamma e^+e^-$ and $\omega\rightarrow \pi^0e^+e^-$; in addition to the background of $\bar c c\rightarrow e^+e^-$. In this work we will focus on the first two decays as Dalitz decays result to be much more complicated to simulate. However, in the left panel of Fig. \ref{accept}, we show how the $\omega$ Dalitz decay is affected by acceptance correction, described below. Work on the incorporation of anomalous propagators in Dalitz decays is in progress. 

From the technical point of view, the acceptance correction constitutes an important issue in order to compare with experimental results. Following these lines, we draw our attention to the PHENIX experiment where a (single lepton) transverse momentum cut is set to $|\vec p^e_t|>200$ MeV and a dilepton rapidity one to $|y_{ee}|<0.35$ so that the outgoing lepton pairs are sufficiently separated to reconstruct such events. Moreover, the lepton pair invariant mass cannot be perfectly resolved as events are collected in bins of 10 MeV. To overcome this limitation, simulations need to incorporate a gaussian smearing of width $\Delta$ of the same size as the bins used experimentally. In Fig. \ref{accept} we show how different channels are affected by acceptance correction.

\vspace{0.5em}

All in all, acceptance correction breaks Lorentz invariance and the final phase space integration cannot be analytically computed but needs of numerical simulation since the easiest process in the hadron cocktail unavoidably requires at least 5 non-trivial integrals. For Dalitz decays, calculations are even more tricky with 7-8 integrals to solve. Therefore, we have used the VEGAS routine to numerically compute the phase space integrals.

\vspace{0.5em}

Finally, the vector mesons contribution to the dilepton productions  after acceptance corrections is given by the following expression:
\begin{align}
\nonumber \frac{dN}{d^4xdM}=&\int d\tilde M\frac{1}{\sqrt{2\pi}\Delta}\exp\left [-\frac{(M-\tilde M)^2}{2\Delta^2}\right ]c_V\frac{\alpha^2}
{24\pi \tilde M}\left (1-\frac{n_V^2 m_\pi^2}{\tilde M^2}\right)^{3/2}\\
\nonumber \times& \sum_{\epsilon=L,\pm}\int_{\text{acc.}}\frac{k_tdk_t dyd^2\vec p_t}
{|E_kp_\parallel-k_\parallel E_p|}\frac{1}{e^{\tilde M_t/T}-1}P_\epsilon^{\mu\nu}\left (\tilde M^2g_{\mu\nu}+4p_\mu p_\nu\right ) \dfrac{m_{V,\epsilon}^4\left (1+\frac{\Gamma_V^2}{m_V^2}\right )}{\left (\tilde M^2-m_{V,\epsilon}^2\right )^2+ m_{V,\epsilon}^4\frac{\Gamma_V^2}{m_V^2}},
\end{align}
where $n_V=2,0$ for $\rho$ and $\omega$ cases respectively; and $M>n_Vm_\pi$. The standard definition for transverse mass $M_t^2=M^2+k_t^2$ is used separating transverse components $\vec k_t$ from parallel ones $k_\parallel$ and the chiral projectors $P_\epsilon^{\mu\nu}$ are detailed in ref. \cite{abnorm}. The coefficient $c_V$ absorbs combinatorial factors different for both channels, meson chemical potential $\mu_V$ and finite volume suppression so as they are not known with precision, the relative weights are used as free parameters in the cocktail by PHENIX collaboration and we shall follow the same procedure here.

\section{Numerical results for dilepton excess}

\begin{figure}[h!]
\centering
\includegraphics[scale=0.25]{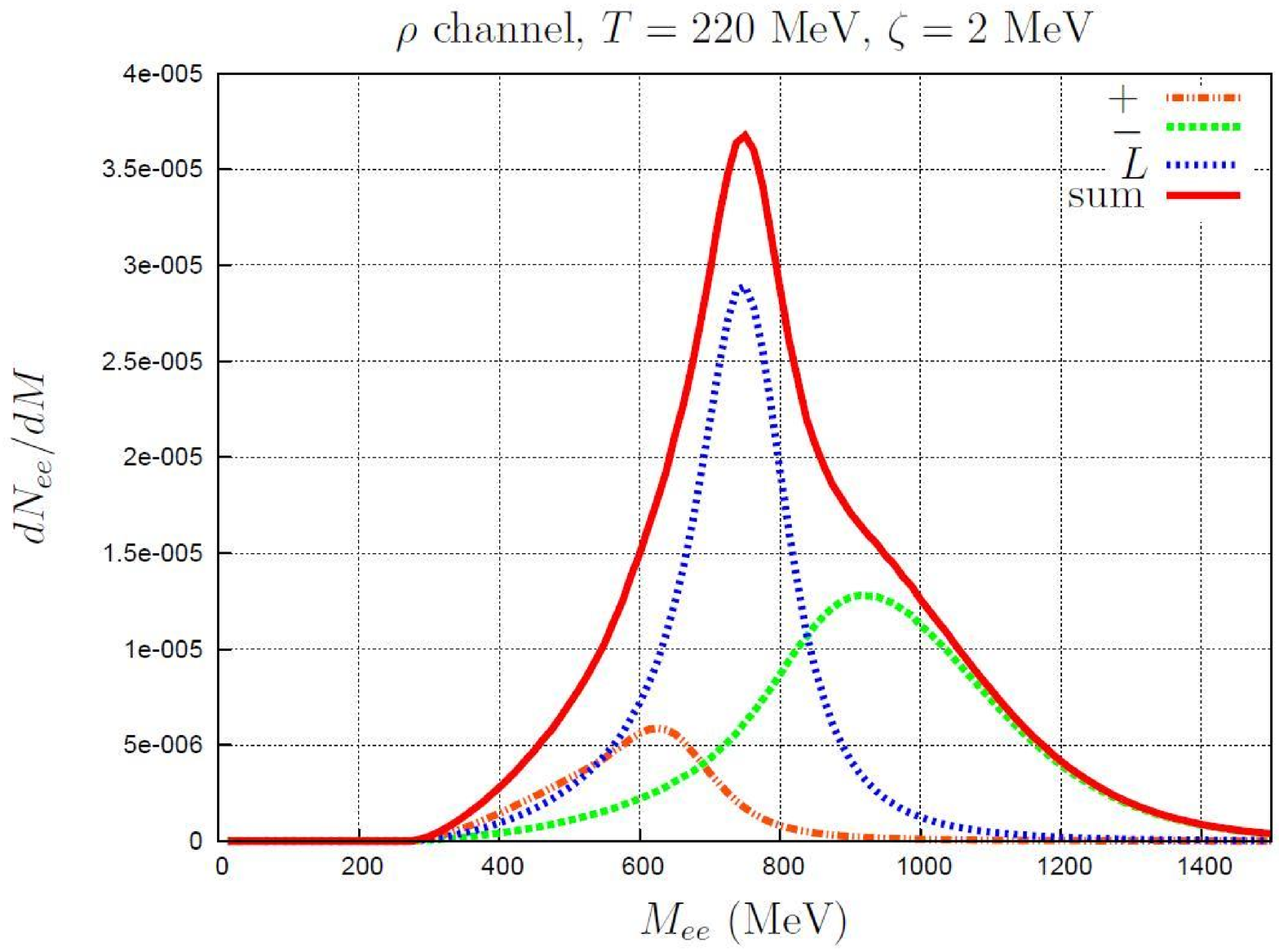}\includegraphics[scale=0.25]{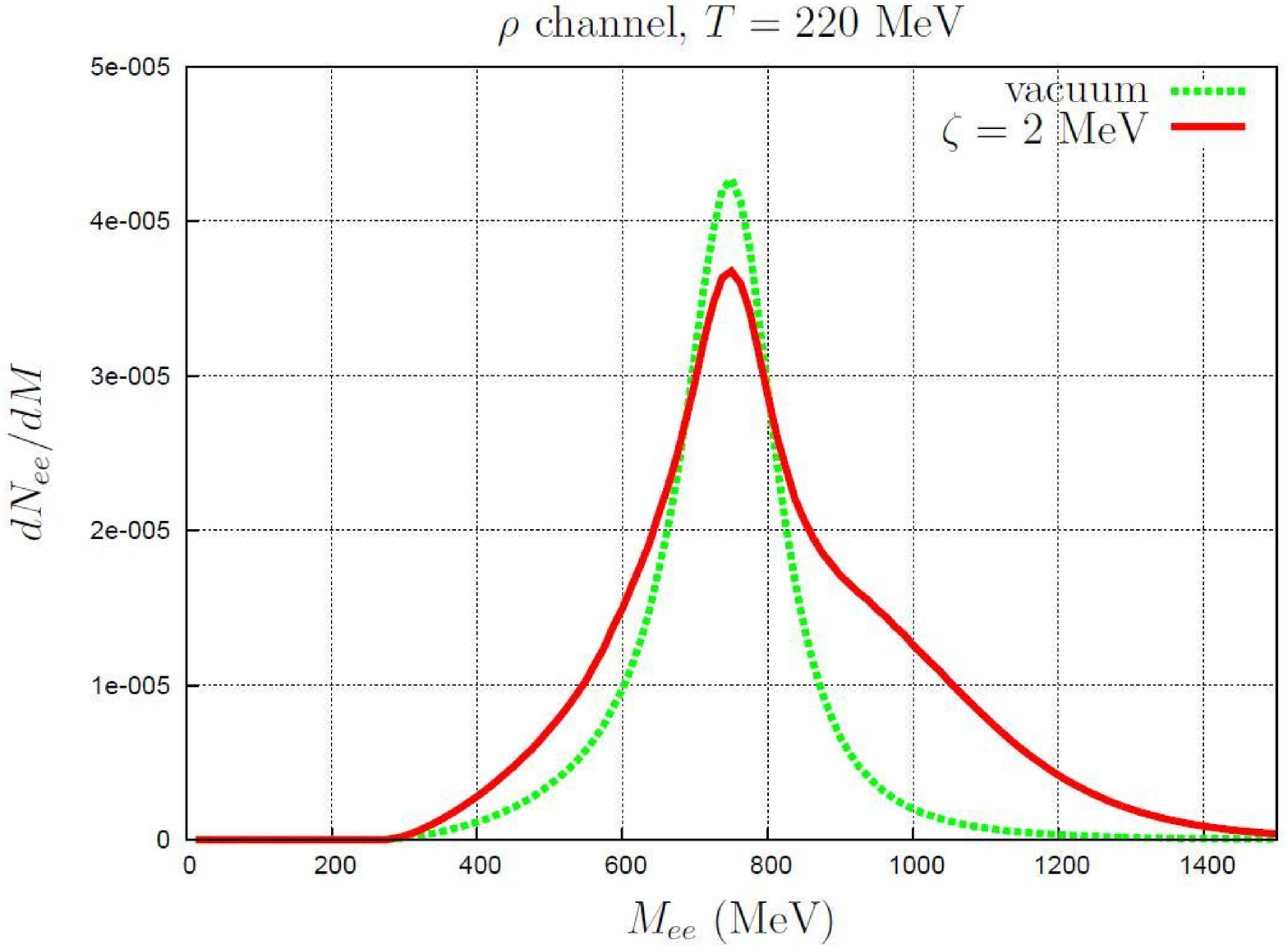}
\vspace{-2em}
\caption{Left: Polarization splitting in $\rho$ spectral function for LPB $\zeta=2$ MeV. Right: Comparison $\rho$ spectral function in vacuum and for LPB $\zeta=2$ MeV.}\label{rhospect}
\end{figure}

Introducing now the characteristic inputs of the PHENIX experiment, namely, the experimental acceptance cuts and the average temperature, that is $T\simeq 220$ MeV, as stated in ref. \cite{phenix}, we present in Fig. \ref{rhospect} (left), the $\rho$ spectral function together with the separate contributions for each different polarization for $\zeta=2$ MeV. Similar results may be obtained modifying this parameter, so when $\zeta$ increases, the chiral resonances appear to be more separated from the vacuum one.

\vspace{0.5em}

Yet, this realistic value allows us to qualitatively show and stress the \textbf{polarization asymmetry} that is perfectly visible aside the peak. In the invariant mass region above the vacuum peak, it is clear that a large percentage of lepton pairs are chiral, unlike expected in vacuum or in medium parity-even calculations. Of course, in the mass region below the peak, the asymmetry is also present but does not show such a spectacular behaviour as above the peak. Therefore, the precise experimental measure of event-by-event dilepton polarization may reveal in an unambiguous way the existence of LPB, confirming the hypothesis of pseudoscalar condensate formation in the fireball of a HIC.
%


\begin{figure}[h!]
\centering
\includegraphics[scale=0.25]{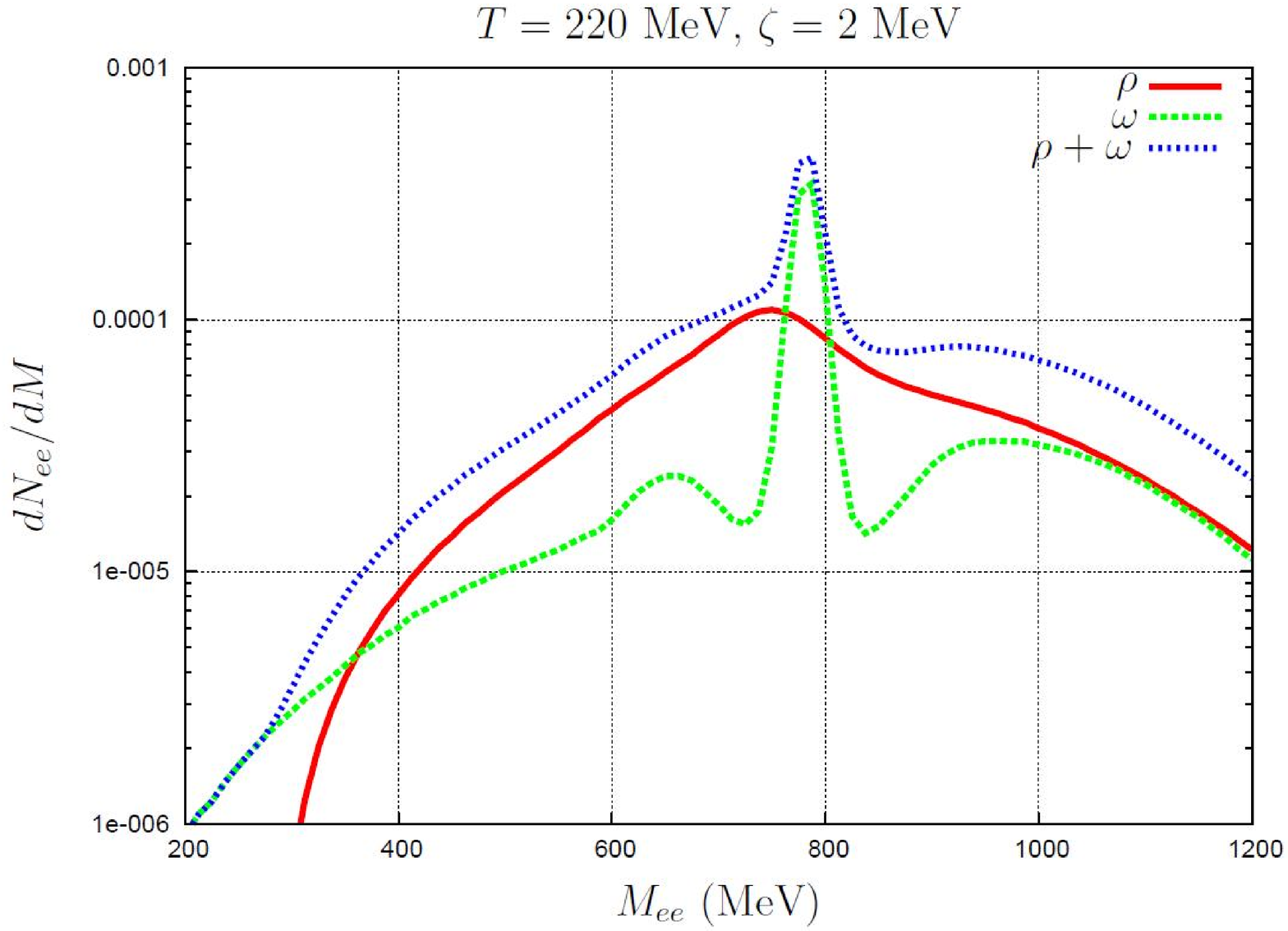}\includegraphics[scale=0.25]{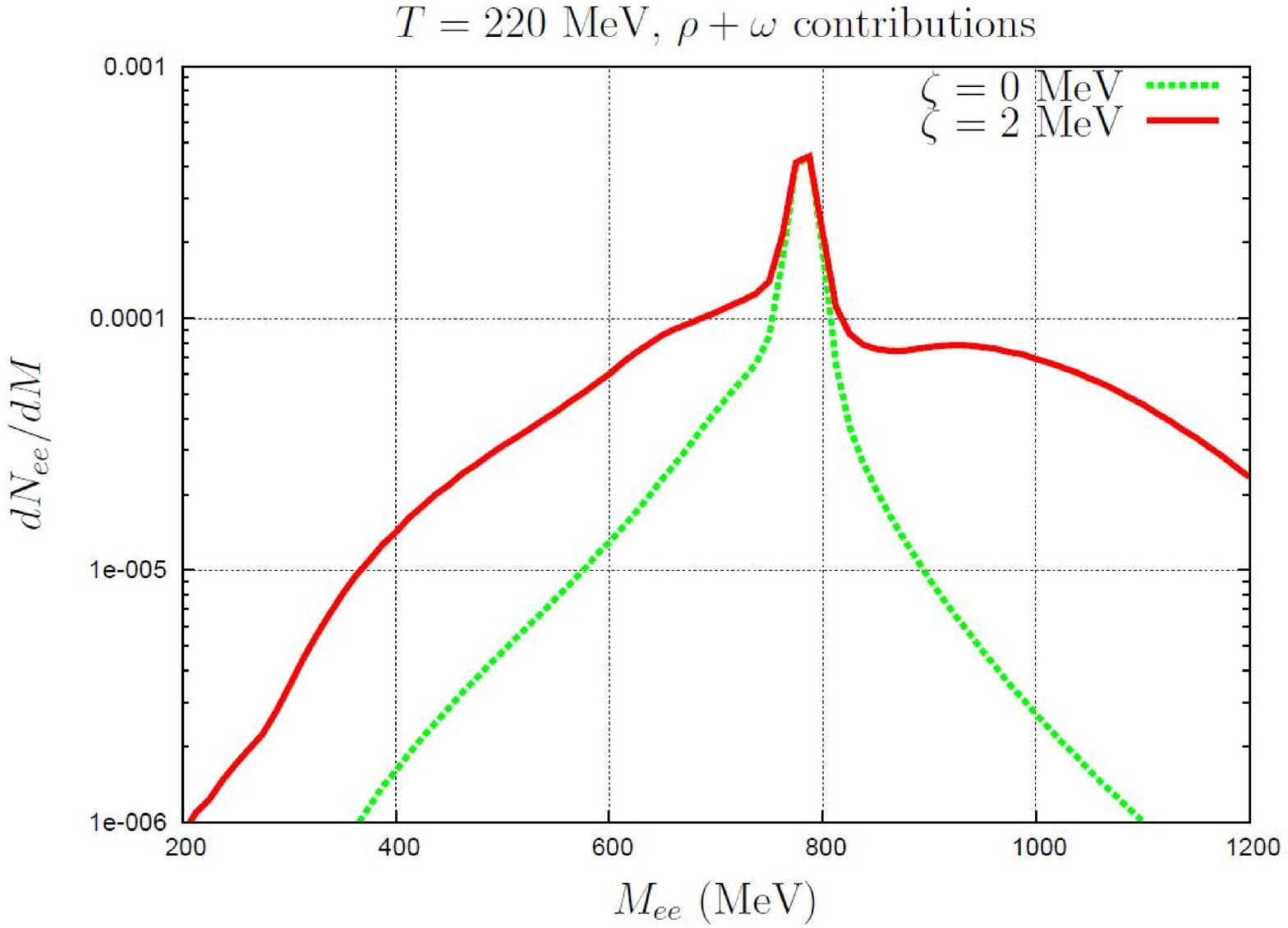}
\vspace{-2em}
\caption{Left: $\rho$ and $\omega$ contributions to dilepton yield for LPB $\zeta=2$ MeV. Right: $\rho$ + $\omega$ contributions in vacuum and for LPB $\zeta=2$ MeV (normalization given by the $\omega$ peak).}\label{rho+w}
\end{figure}



\begin{figure}[h!]
\begin{minipage}[b]{3.4cm}
\centering
\includegraphics[scale=0.24]{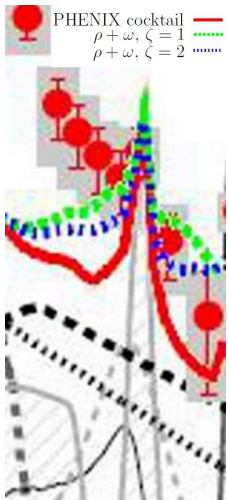}
\vspace{-1em}
\caption{Comparison of PHENIX cocktail (red) with modified cocktail using $\rho$ + $\omega$ contributions for LPB with $\zeta=$1 (green) and 2 MeV (blue).\label{cockt}}

\end{minipage}\qquad \begin{minipage}[b]{10.95cm}

\qquad In Fig. \ref{rhospect} (right), we also show the comparison of the $\rho$ spectral function in vacuum and the one for LPB with $\zeta=2$ MeV. A remarkable point is the fact that in a HIC, unlike in vacuum, thermal equilibrium may be reached and a pion-pion recombination into $\rho$ may easily take place, so the effective number of $\rho$ mesons within the fireball becomes larger than the one expected as a direct result from the very collision (about a factor of 1.8, according to \cite{phenix}).\vspace{0.5em}

\qquad In addition, we show in Fig. \ref{rho+w} the separate contribution of $\rho$ and $\omega$ for $\zeta=2$ MeV and their sum (left); and in the right panel, we compare this sum with the one corresponding to the vacuum contribution with a normalization given by the $\omega$ peak.\vspace{0.5em}

\qquad Finally, in Fig. \ref{cockt} we present the PHENIX cocktail after subtraction of $\rho$ and $\omega$ channels and addition of our modified spectral functions for LPB with $\zeta=$1, 2 MeV. Indeed, an enhancement of the dilepton yield aside the resonance arises due to parity-breaking effects and may certainly help to explain the PHENIX puzzle. However, the region $M<700$ MeV has not been accounted yet and the incorporation of Dalitz $\omega$ and $\eta$ decays is also expected to provide an excess of lepton pairs.

\end{minipage}
\end{figure}


\section{Conclusions}

As discussed, LPB is not forbidden by any physical principle in QCD at finite temperature and density so the possibility of a pseudoscalar condensate to appear as a result from a HIC inside the hot nuclear fireball has been discussed. The effect leads to unexpected isospin-dependent modifications of vector mesons in-medium properties. After a theoretical analysis, one may to compute the dilepton production in the PHENIX experiment, thus realizing that LPB seems capable of explaining in a natural way the abnormal dilepton production. Moreover, we also want to stress that event-by-event measurements of the lepton polarization asymmetry aside the $\omega$ peak may reveal in an ambiguous way the existence of LPB.

\vspace{0.5em}

Finally, the Dalitz $\omega$ and $\eta$ decays are thought to be mainly responsible for the enhancement at $300<M<700$ MeV, while the isotriplet condensate could also play an important role and should be analysed and incorporated to the calculation. This work is in progress and we expect it to be published soon.

\end{document}